# Thermal radiation and blackbody radiation drag of a large-sized perfectly black particle moving with relativistic velocity

A.A. Kyasov and G.V. Dedkov

Nanoscale Physics Group, Kabardino-Balkarian State University, Nalchik, 360000, Russia

We have developed a self-consistent description of the radiation heat transfer and dynamics of large perfectly black spherical bodies with sizes much greater than the characteristic wavelength of radiation   moving in a photon gas with relativistic velocity. The results can be important in astrophysics.

**1. Introduction**

In 1968, several authors have calculated the force acting on a perfectly black spherical particle from the point of view of an observer in arbitrary uniform motion with respect to the c.m. of the radiation [1] (see also references therein), i. e. in the frame of reference co-moving with particle. However, the corresponding expression for this force in the reference frame of the blackbody radiation was not obtained. Important questions concerning the dynamics and intensity of emitted and absorbed thermal radiation of particle in the reference frame of the blackbody radiation were also not addressed.

At the same time, the questions of dynamics, thermal radiation and heating (cooling) of a large-sized body are closely interrelated. Therefore, a comprehensive solution to this problem requires calculating the intensity of thermal radiation and radiation force both in the reference frame of radiation and in the reference frame co-moving with the body.

For a dipole-like dielectric particle in the long-wavelength approximation $R << \lambda$ ( $R$ and $\lambda$ are the particle radius and the characteristic wavelength of radiation) all this has been done in [2]. However, for a perfectly black particle in the short-wavelength approximation $R >> \lambda$ (geometrical optics approximation), closed self-consistent solution to this problem is still absent. This work aims at filling this gap, and the results provide the link between the limits $R << \lambda$ and $R >> \lambda$ , forming closed physical picture.

## 2. Theory

Figure 1 shows the problem statement and the frames of reference $\Sigma$, $\Sigma'$ used in what follows. $\Sigma$ is the frame of blackbody radiation with temperature $T_2$, and $\Sigma'$ is the rest frame of a particle. In the $\Sigma$-frame a particle with temperature $T_1$ moves in the direction of the positive $x$-axis with velocity $V = \beta c$ ($c$ is the speed of light in vacuum).

We assume the geometrical optics approximation to be fulfilled (short-wavelength approximation)

$$R \gg \max\left(\frac{2\pi \hbar c}{k_B T_1}, \frac{2\pi \hbar c}{k_B T_2}\right) \tag{1}$$

Then, according to [1], the dissipative force acting on the particle in $\Sigma'$, is given by

$$F'_x = -\frac{4}{3}\frac{\beta \gamma^2}{c} a T_2^4 \tag{2}$$

where $\gamma = \frac{1}{\sqrt{1-\beta^2}}$, $a = 4\pi R^2 \sigma_B$, $\sigma_B = \frac{\pi^2}{60}\frac{k_B^4}{\hbar^3 c^2}$ is the Stefan-Boltzmann constant. Moreover, the energy density of an equilibrium electromagnetic field (photon gas) is given by [1]

$$\varepsilon' = \frac{4}{c}\sigma_B T_2^4 \gamma^2 \left(1 + \frac{\beta^2}{3}\right) \tag{3}$$

The validity of (2), (3) is easily verified using the energy-momentum tensor of the equilibrium electromagnetic field (written in $\Sigma$) [3]

$$T_{\mu\nu} = (p + \varepsilon)u_\mu u_\nu - p g_{\mu\nu} \tag{4}$$

where $p = \varepsilon/3$, $\varepsilon = \frac{4}{c}\sigma_B T_2^4$; $\mu,\nu = 0,1,2,3$. By transforming $T_{\mu\nu}$ into $\Sigma'$, the $F'_x$ and $\varepsilon'$ are expressed through the components of $T'_{\mu\nu}$ [1].

The following analysis is based to great extent on the relationships [2]

$$\dot{Q}' = \gamma^2 \dot{Q} \tag{5}$$

$$I \equiv I_1 - I_2 = -\left(\frac{dQ}{dt} + F_x V\right) \tag{6}$$

$$F'_x = F_x - \beta \gamma^2 \dot{Q}/c \tag{7}$$

where $I_1$ and $I_2$ are the intensities of emitted and absorbed radiation in $\Sigma$; $\dot{Q}$ and $\dot{Q}'$ are the rates of the particle heat exchange in $\Sigma$ and $\Sigma'$; $F_x$ and $F_x'$ are the forces acting on the particle in $\Sigma$ and $\Sigma'$, respectively. Originally, equations (5)—(7) were obtained using the long-wavelength approximation $R \ll \lambda$ which is opposite to (1) [2]. However, as shown in Appendix, they are valid also in the case $R \gg \lambda$. It is worth noting that the equations similar to (6), (7) were earlier obtained by Polevoi [4] in the problem of vacuum friction between the two semiinfinite half-spaces in relative motion. In general case, Eqs. (6), (7) appear due to the condition of quasistationarity and their validity is independent of the relation between $R$ and $\lambda$.

Further calculations are very simple. According to the Stefan-Boltzmann law, the intensity of thermal radiation in $\Sigma'$ is given by

$$I_1' = \sigma_B T_1^4 4\pi R^2 = aT_1^4 \tag{8}$$

The intensity of absorbed radiation in $\Sigma'$ is obtained using the relation between $I_2'$ and $\varepsilon'$:

$$I_2' = \frac{c}{4}\varepsilon' 4\pi R^2 \tag{9}$$

Using (9) and (3) yields

$$I_2' = aT_2^4 \gamma^2 \left(1 + \frac{\beta^2}{3}\right) \tag{10}$$

On the other hand, the rate of particle heating (cooling) in $\Sigma'$ takes the form

$$\frac{dQ'}{dt'} = \dot{Q}' = I_2' - I_1' = aT_2^4 \gamma^2 \left(1 + \frac{\beta^2}{3}\right) - aT_1^4 \tag{11}$$

Using (5) and (11) one obtains the equation of heating (cooling) in $\Sigma$

$$\frac{dQ}{dt} = \dot{Q} = aT_2^4 \left(1 + \frac{\beta^2}{3}\right) - \frac{1}{\gamma^2} aT_1^4 \tag{12}$$

Moreover, by inserting (2) and (12) into (7) we obtain the radiation force $F_x$

$$F_x = -\frac{\beta}{c} a\left(T_1^4 + \frac{1}{3}T_2^4\right) \tag{13}$$

From (13) it follows that, in contrast to $F_x'$, radiation force $F_x$ in $\Sigma$ contains the contribution both vacuum radiation and the contribution of radiation emitted by the particle. The same situation takes place for dielectric particles in the long-wavelength approximation [2]. Finally, by inserting (12) and (13) into (6) one obtains

$$I = a(T_1^4 - T_2^4) \tag{14}$$

Surprisingly, Eq. (14) does not depend on the particle velocity and reduces formally to the static result. However, paradoxicalness of this fact vanishes if we take into account that as a result of heat exchange with the background radiation the particle rather quickly reaches a steady-state temperature (assuming that $T_2 = const$), defined by the condition (see the next Section)

$$\dot{Q}' = \dot{Q} = 0 \tag{15}$$

From (12) and (15) the steady-state temperature of a moving perfectly black body is given by

$$T_s = T_2 \gamma^{1/2} (1 + \beta^2 / 3)^{1/4} \tag{16}$$

Inserting (16) into (14) yields the stationary imbalance of radiation and absorption

$$I_s = aT_2^4 \left[ \gamma^2 (1 + \beta^2 / 3) - 1 \right] \tag{17}$$

Both $T_s$ and $I_s$ depend on the particle velocity, as it could be expected. Formulas (12)–(14) and (16), (17) are the main results of this paper. In the case when a large-sized particle is not perfectly black and is characterized by the absorptivity $a_a$ ($0 < a_a < 1$), and emissivity $a_r$ ($0 < a_r < 1$) these formulas can be trivially modified. In the case when the particle is characterized by the dielectic and magnetic properties, Eqs. (5)–(7) have to be used along with Eqs. (A7) and (A13) (see Appendix).

## 3. Kinetics of heating/cooling and dynamics of particle

It is interesting to compare the time-scale of particle deceleration and the time needed to reach the steady-state temperature. The dynamics equation (in $\Sigma$) has the form

$$\frac{d}{dt}\left(\frac{mV}{\sqrt{1-\beta^2}}\right) = F_x \tag{18}$$

Along with the temporal velocity dependence in the left-hand side of (17), we must take into account the change in the particle mass due to the radiation and absorption. With allowance for this fact Eq. (18) takes the form [2]

$$m\frac{dV}{dt} = \gamma^{-3} F'_x \qquad (19)$$

Substituting (2) into (19) yields

$$\frac{d\beta}{\beta\sqrt{1-\beta^2}} = -\frac{4aT_2^4}{3mc^2} dt \qquad (20)$$

After integrating (20) one obtains

$$\frac{\gamma(t)-1}{\gamma(t)+1} = \frac{\gamma(0)-1}{\gamma(0)+1} \exp\left(-\frac{8aT_2^4}{3mc^2} t\right) \qquad (21)$$

where $\gamma(t) = \left(1-V(t)^2/c^2\right)^{-1/2}$, $\gamma(0) = \left(1-V(0)^2/c^2\right)^{-1/2}$, $V(0)$ and $V(t)$ are the velocities at the moments $t = 0$ and $t$. From (21), the characteristic deceleration time is given by

$$\tau_V = \frac{3mc^2}{8aT_2^4} = \frac{R\rho c^2}{8\sigma_B T_2^4} \qquad (22)$$

where $\rho$ is the particle density. For example, the time $\tau_V$ for an icy H$_2$O particle at $R = 20\,cm$, $\rho \approx 1\,g/cm^3$ and $T_2 = 100\,K$ is $\tau_V \approx 12 \cdot 10^9$ years, i. e. close to the age of the Universe.

To analyze the kinetics of heating, we represent $\dot{Q}'$ in the form

$$\frac{dQ'}{dt'} = \frac{d}{dt'}(C_s m T_1) = \gamma C_s m \frac{dT_1}{dt} + \gamma C_s T_1 \frac{dm}{dt} + \gamma m T_1 \frac{dC_s}{dt} \qquad (23)$$

With allowance for the relations $dm/dt = \gamma \dot{Q}/c^2$ from (5) and (23) one obtains

$$C_s m \frac{dT_1}{dt} = \gamma \dot{Q}(1 - C_s T_1/c^2) - m T_1 \frac{dC_s}{dt} \qquad (24)$$

Since $C_s T_1/c^2 \ll 1$ at any possible temperature of solids, the corresponding term in (24) can be omitted. Assuming $C_s = const$, substituting (12) into (24) and introducing the reduced variables of time $\tau = t/\tau_Q$ (with $\tau_Q = \dfrac{C_s \rho R}{3\sigma_B T_2^3}$), and temperature $x = T_1/T_2$, Eq. (24) takes the form

$$\frac{dx}{d\tau} = \gamma(1 + \beta^2/3) - \frac{x^4}{\gamma} \qquad (25)$$

As follows from the numerical solution to (25), at any initial value of $x$ it goes to the steady-state value $x_s = T_s/T_2$ (provided that the background temperature $T_2$ is constant). The characteristic dimensionless time of this process is of order 1 and decreases with increasing $\gamma$. Therefore, the time scale of particle heating/cooling is determined by the parameter $\tau_Q$. In the case $\gamma \gg 1$, the third term in (25) is negligible, and the steady-state temperature is given by (16), in accordance with (15). However, at $\gamma \sim 1$ this contribution becomes important, and the steady-state temperature and radiation depend on the function $C_s(T_1)$.

Comparing $\tau_Q$ and $\tau_V$ yields $\tau_Q/\tau_V = \dfrac{8 C_s T_2}{3c^2} \approx 10^{-10} \div 10^{-14}$ at typical values $C_s = 10 \div 10^3 \, J/K \cdot kg$ and $T_2 = 10 \div 10^3 \, K$. Therefore, the process of heating/cooling proceeds much faster than the deceleration.

## 4. Conclusions

A self-consistent description of radiation heat transfer and dynamics of large particles moving in a photon gas with relativistic velocity has been developed. The radiation force acting on a particle in the reference frame of background radiation is found. Initially, the difference between the emission and absorption does not depend on the particle velocity, and it is described by the static expression via the Stefan-Boltzmann law. It is shown that the process of heating/cooling proceeds much faster than the process of deceleration and the particle acquires a steady-state temperature proportional to the temperature of backround radiation. In this case, the intensity of thermal radiation is proportional to the gamma-factor squared being considerably higher than the intensity of absorbed radiation. Further on, the particle is gradually slowing down and its kinetic energy is converted into radiation. With slight modification, these results can be applied to not perfectly black bodies. For the bodies with well defined dielectric and magnetic properties the results can be obtained numerically.

Apart from the general theoretical meaning, the results can be used in studying the evolution of cosmic bodies and their thermal radiation. Rather typical velocities of cosmic bodies may reach $10^3$–$10^4$ km/s. Though in the state of thermal equilibrium of the system consisting of solid bodies and radiation the difference between emission and absorption decreases, this difference may be an indicator of the internal dynamics for an external observer.

**Appendix**

Here we recall the known relativistic transformations of the electric density current *j*, electric and magnetic fields **E**, **B**, polarization **P** and magnetization **M** corresponding to the Lorentz transformations from $\Sigma'$ to $\Sigma$:

$$j'_x = \gamma(j_x - V\rho); \quad j'_y = j_y; \quad j'_z = j_z \tag{A1}$$

$$E'_x = E_x; \quad E'_y = \gamma(E_y - \beta B_z); \quad E'_z = \gamma(E_z + \beta B_y) \tag{A2}$$

$$B'_x = B_x; \quad B'_y = \gamma(B_y + \beta E_z); \quad B'_z = \gamma(B_z - \beta E_y) \tag{A3}$$

$$P'_x = P_x; \quad P'_y = \gamma(P_y + \beta M_z); \quad P'_z = \gamma(P_z - \beta M_y) \tag{A4}$$

$$M'_x = M_x; \quad M'_y = \gamma(M_y - \beta P_z); \quad M'_z = \gamma(M_z + \beta P_y) \tag{A5}$$

Consider the Joule dissipation integral in $\Sigma'$

$$\int \langle \mathbf{j}'\mathbf{E}' \rangle d^3 r' = \int \langle j'_x E'_x + j'_y E'_y + j'_z E'_z \rangle d^3 r', \tag{A6}$$

where the angular brackets $\langle ... \rangle$ imply the total quantum-statistical averaging. In (A6), the integration is performed over the volume of the particle and we extend it to the whole space. This is assumed in what follows for all integrals. Substituting (A1), (A2) into (A6) yields [5]

$$\int \langle \mathbf{j}\mathbf{E} \rangle d^3 r = F_x V + \gamma^{-2} \int \langle \mathbf{j}'\mathbf{E}' \rangle d^3 r', \tag{A7}$$

where $F_x$ is the *x*-component of the Lorentz force

$$\mathbf{F} = \int \langle \rho \mathbf{E} \rangle d^3 r + \frac{1}{c} \int \langle \mathbf{j} \times \mathbf{B} \rangle d^3 r \tag{A8}$$

Taking the density current in $\Sigma'$

$$\mathbf{j}' = \frac{\partial \mathbf{P}'}{\partial t'} + c \operatorname{rot} \mathbf{M}' \tag{A9}$$

and substituting (A9) into (A6) we get

$$\int \langle \mathbf{j}' \mathbf{E}' \rangle d^3 r' = \int \left\langle \left( \frac{\partial \mathbf{P}'}{\partial t'} + c \operatorname{rot} \mathbf{M}' \right) \mathbf{E}' \right\rangle d^3 r' =$$

$$= \int \left\langle \frac{\partial \mathbf{P}'}{\partial t'} \mathbf{E}' + c(\mathbf{E}' \operatorname{rot} \mathbf{M}' - \mathbf{M}' \operatorname{rot} \mathbf{E}' + \mathbf{M}' \operatorname{rot} \mathbf{E}') \right\rangle d^3 r' =$$

$$= \int \left\langle \frac{\partial \mathbf{P}'}{\partial t'} \mathbf{E}' + c \operatorname{div}(\mathbf{M}' \times \mathbf{E}') + c \mathbf{M}' \operatorname{rot} \mathbf{E}' \right\rangle d^3 r' = \qquad (A10)$$

$$= c \oint (\mathbf{M}' \times \mathbf{E}') d\mathbf{S}' + \int \left\langle \frac{\partial \mathbf{P}'}{\partial t'} \mathbf{E}' - \mathbf{M}' \frac{\partial \mathbf{B}'}{\partial t'} \right\rangle d^3 r' =$$

$$= c \oint (\mathbf{M}' \times \mathbf{E}') d\mathbf{S}' + \int \left\langle \frac{\partial \mathbf{P}'}{\partial t'} \mathbf{E}' + \frac{\partial \mathbf{M}'}{\partial t'} \mathbf{B}' \right\rangle d^3 r' - \int \left\langle \frac{\partial}{\partial t'} (\mathbf{M}' \mathbf{B}') \right\rangle$$

In the last line of (A10), the integral over infinitely remote surface equals zero, while the third term is also zero due to the stationarity of electromagnetic fluctuations. Then it follows from (A10)

$$\dot{Q}' = \int \langle \mathbf{j}' \mathbf{E}' \rangle d^3 r' = \int \left\langle \frac{\partial \mathbf{P}'}{\partial t'} \mathbf{E}' + \frac{\partial \mathbf{M}'}{\partial t'} \mathbf{B}' \right\rangle d^3 r' \qquad (A11)$$

Substituting (A2)–(A5) into (A11) and making use simple transformations with allowance for $dt' = \gamma^{-1} dt$, $d^3 r = \gamma^{-1} d^3 r'$ one obtains

$$\int \left\langle \frac{\partial \mathbf{P}'}{\partial t'} \mathbf{E}' + \frac{\partial \mathbf{M}'}{\partial t'} \mathbf{B}' \right\rangle d^3 r' = \gamma^2 \int \left\langle \frac{\partial \mathbf{P}}{\partial t} \mathbf{E} + \frac{\partial \mathbf{M}}{\partial t} \mathbf{B} \right\rangle d^3 r \qquad (A12)$$

Therefore, we proved that the variable

$$\dot{Q} = \int \langle \dot{\mathbf{P}} \mathbf{E} + \dot{\mathbf{M}} \mathbf{B} \rangle d^3 r \qquad (A13)$$

is transformed from $\Sigma$ to $\Sigma'$ according to the Eq. (5):

$$\dot{Q}' = \gamma^2 \dot{Q} \qquad (A14)$$

Since we did not use any restrictions on the value of radius, Eq. (A14) is valid irrespectively of the relation between the particle radius $R$ and the characteristic wavelength of radiation $\lambda$. As already was said, the origin of Eqs. (6) and (7) is due to the condition of quasistationarity and therefore Eqs. (5)–(7) have the general character.

It is also worth noting that using (A13), (A14), Eq. (A7) takes the form

$$\int \langle \mathbf{j} \mathbf{E} \rangle d^3 r = \left( \int \langle \rho \mathbf{E} \rangle d^3 r + \frac{1}{c} \int \langle \mathbf{j} \times \mathbf{B} \rangle d^3 r \right) \cdot \mathbf{V} + \int \langle \dot{\mathbf{P}} \mathbf{E} + \dot{\mathbf{M}} \mathbf{B} \rangle d^3 r \qquad (A15)$$

Eq. (A15) generalizes the analogous equation for a small particle with fluctuatingdipole moments $\mathbf{d}$ and $\mathbf{m}$ [2, 5]:

$$\int_{(V)} \langle \mathbf{jE} \rangle d^3r = \left( \langle \nabla(\mathbf{dE}+\mathbf{mB}) \rangle \right) \cdot \mathbf{V} + \langle \mathbf{\dot{d}E}+\mathbf{\dot{m}B} \rangle \tag{A16}$$

However, Eq. (A16) is valid under the condition $R \ll \lambda$.

Figure 1

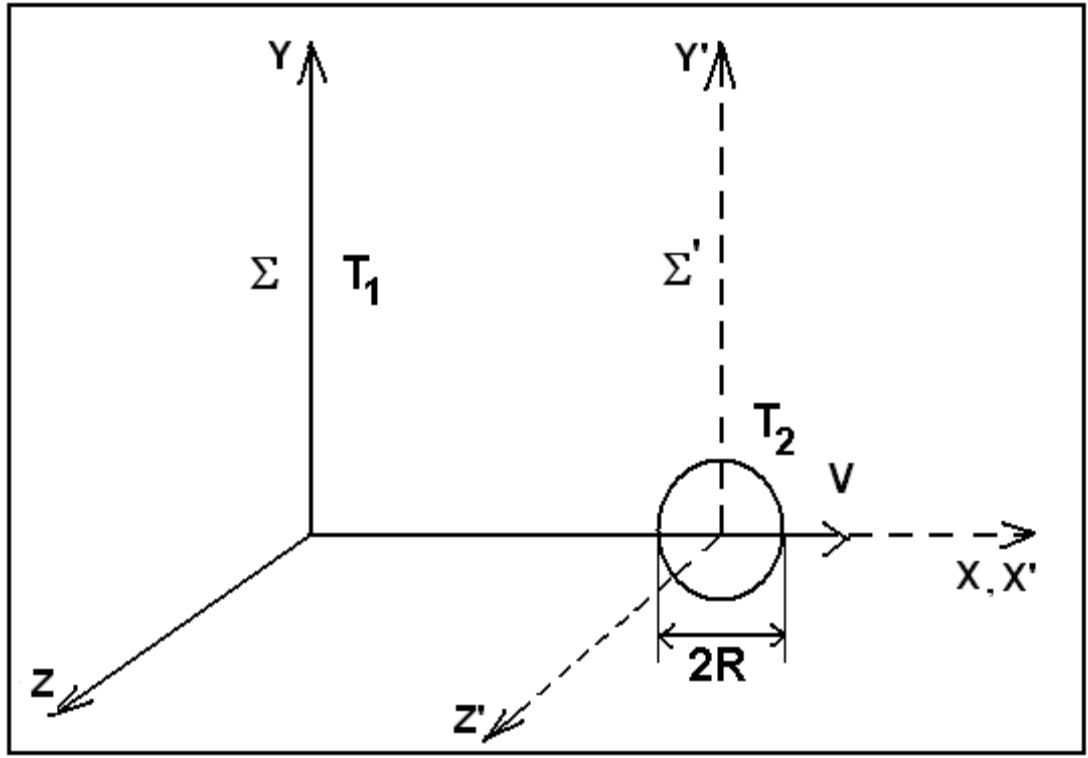

Geometrical configuration and reference frames corresponding to the vacuum background ($\Sigma$) and particle ($\Sigma'$). $\Sigma'$ moves along $+x'$ with velocity $V$.